\documentclass[a4paper,12pt]{article}
\usepackage{latexsym}
\newcommand\emt{energy--momentum tensor}
\newcommand\lag{Lagrangian}

\newcommand\Glin{G^{\rm L}}
\newcommand\GR{General Relativity}
\begin{document}
\title{Symmetry properties\\ under arbitrary field redefinitions\\
of the metric energy--momentum tensor\\ in classical field theories and
gravity}
\author{Guido MAGNANO \\ Dipartimento di Matematica,
Universit\`a di Torino, \\ via Carlo Alberto 10,
10123 Torino, Italy \\ and \\
Leszek M. SOKO\L{}OWSKI \\ Astronomical Observatory,
Jagellonian University, Orla
171, \\ Krak\'ow 30-244, Poland}
\date{}
\maketitle
\begin{abstract}
We derive a generic identity which holds for the metric
(i.e.~variational)
   energy--momentum
tensor under any field transformation in any generally covariant
classical
   Lagrangian field
theory. The identity determines the conditions under which a symmetry of
the
Lagrangian is
also a symmetry of the energy--momentum tensor. It turns out that the
stress
tensor acquires
the symmetry if the Lagrangian has the symmetry in a generic curved
spacetime.
In this sense a
field theory in flat spacetime is not self--contained. When the identity
is
applied to the
gauge invariant spin-two field in Minkowski space, we obtain an
alternative
and direct
derivation of a known no--go theorem: a linear gauge invariant spin-2
field,
which is
dynamically equivalent to linearized General Relativity, cannot have a
gauge
invariant metric
energy--momentum tensor. This implies that attempts to define the notion
of
gravitational
energy density in terms of the metric energy--momentum tensor in a
field--theoretical formulation of gravity
must fail.
\end{abstract}

\section{Introduction}

Total energy and energy density are clearly among the most significant
physical
quantities in any field theory. The lesson we have learnt from
Einstein's
\GR\
is that the adequate description of energy and momentum of any kind of
matter
and field,
except for the gravitational field itself, is in terms of the
variational
(with respect to the
spacetime metric) \emt (the metric stress tensor, for short).

In the gauge theories of particle physics the metric stress tensors for
the
gauge
fields are all gauge invariant. This may arouse a conviction that
this is a generic feature of
any gauge invariant theory. However this is not the case. In general
the metric stress
tensor does not
inherit the gauge--independence property of the underlying
Lagrangian. The most important
example is linearized \GR. This theory is dynamically equivalent to
the linear massless
spin--two field, whose metric stress tensor is gauge dependent
\cite{AD2}.  Then two
related problems arise. Why does the metric stress tensor for a gauge
invariant field in Minkowski space lose this symmetry? Is it possible to
construct a linear spin--two field theory in flat spacetime with a
gauge invariant metric
stress tensor?

An answer to the first problem follows from a "folk theorem",
rigorously stated and proven by
Deser and McCarthy in \cite{DMC} to the effect that the Poincar\'e
generators, being spatial
integrals of the metric stress tensor, are gauge invariant and thus
unique. The
theorem shows that in quantum field theory, where only global (i.e.
integrals
over all 3--space) quantities, such as total energy and momentum of a
quantum
system, are physical (measurable) ones, the inevitable gauge dependence of
the
metric stress tensor (for fields carrying spin larger than one) is quite
harmless. It follows from the proof that the gauge dependence of this
tensor is
due to the fact that the gauge transformations involve the spacetime
metric. The
negative answer to the second problem is also contained in that work: it
is
stated there that this gauge dependence is unavoidable (see also
sect.~4 below).

In a classical gauge invariant field theory any gauge dependence of the
metric stress tensor is truly harmful since this tensor cannot act as
the source
in Einstein field equations and this defect makes the theory
unphysical. Even if
such a field is viewed as a test one in a fixed spacetime, its theory
remains
defective since the local conserved currents (which exist if there are
Killing
vectors) do not determine physical flows of energy or momentum
through a boundary
of a spatially bounded region.

One may consider arbitrary field redefinitions depending on
additional quantities
and some of these may be symmetries of specific Lagrangians. It is
interesting to
see how generic is the case of gauge symmetry breaking by the metric
stress tensor for
fields with spins $s>1$.  To get an overall and unified picture we
investigate in
the present paper
symmetry properties of the stress tensor in any classical
generally covariant Lagrangian field theory. We establish conditions
under which
any symmetry (which is "internal" in the sense that it is not mere
covariance
under coordinate transformations) of the underlying Lagrangian becomes
the
symmetry of the metric stress tensor. To this end we derive a generic
identity,
valid for any Lagrangian in a curved spacetime for an arbitrary field
transformation (being a symmetry or not), which determines how the
stress tensor
is altered under this transformation. This identity is the thrust of the
paper.
For a symmetry transformation of the Lagrangian the identity shows that
the
metric stress tensor inherits the symetry property provided that
either the field
equations hold or the transformation is metric independent. Thus it turns
out
that the spacetime metric plays a key role for all symmetries in a field
theory
and not only for gauge invariance. Furthermore it follows from the
identity that
if a symmetry of the field Lagrangian holds only in flat spacetime (more
generally: in a narrow class of curved spacetimes) and is broken in a
generic
curved one (actually in an open neighbourhood of flat spacetime or of
that class,
respectively), then the symmetry is broken for the metric stress tensor
even in Minkowski space (this class of spacetimes).

As a first application of the identity we provide a
different derivation of the result already given in \cite{DMC} that
all known metric
stress tensors for a linear spin-two field (and for higher spins) in
flat
spacetime are gauge dependent. Next the identity furnishes an alternative
(to
that implicit in \cite{DMC}), simple and direct proof of a no--go
theorem: if in the
weak--field approximation gravity is described by a gauge invariant linear
and
symmetric tensor field and its stress tensor is quadratic and
contains no higher
than second derivatives of the field, then this tensor cannot be
gauge invariant.
If the assumptions are valid, then no meaningful notion of local
gravitational
energy can arise in this way. Since the identity works equally well
for any other
field symmetry we expect that it will find further nontrivial
applications.

       The paper is organized as follows. Section 2 is
the heart of the paper: we derive there the identity for the stress
tensor
variation under any field redefinition. In section 3 we study
the gauge invariant linear massless spin-two field in Minkowski space and
show,
as a direct consequence of the identity, that the metric stress
tensor breaks the gauge symmetry. Following the same way the no--go
theorem for
the weak--field approach to linear gravity is proven in section 4.
Conclusions
are contained in section 5.

\section{Transformation of the metric stress tensor under a field
redefinition}

Let $\phi$ be a dynamical
field or a multiplet of fields (indices suppressed) described by a
generally
covariant action
functional with a Lagrangian density $\sqrt{-g}L(\phi, g_{\mu\nu})$,
residing in a
curved spacetime with a
(dynamical or background) metric $g_{\mu\nu}$; for simplicity we
assume that $L$ does not
depend on second and higher derivatives of
$\phi$. Let $\phi=\varphi(\phi', \xi, g)$ be any invertible
transformation of the dynamical variable, which in general involves
the metric tensor\footnote{Throughout the paper the symbol $g$ has
two distinct meanings:
it denotes either the determinant of the metric, or the metric
itself. The ambiguity
should not cause any confusion. The covariant derivatives
$\nabla_\alpha f\equiv f_{;\alpha}$ are taken
with respect to the spacetime metric $g_{\mu\nu}$.}
and a non--dynamical vector or tensor field $\xi$ and its first
covariant derivative
$\nabla\xi$. The transformation is arbitrary with the exception that
we exclude the tensor
(or spinor) transformations of the field under a mere coordinate
transformation.
As a consequence, as opposed to many authors (see e.g. \cite{A},
\cite{ABM}, \cite{PSS}), we do
not view the transformation group of the dynamical variables induced
by spacetime
diffeomorphisms as a gauge group. The transformation need not be
infinitesimal.
Under the change of the dynamical field one sets
\begin{equation}\label{NL}
L(\phi, g)=L(\varphi(\phi', \xi, g), g)\equiv L'(\phi', \xi, g).
\end{equation}
The variational (metric) stress tensor\footnote{We emphasize
that $L$ is a matter Lagrangian. Whenever the metric is regarded as a
dynamical
field one should add to $L$ a separate gravitational Lagrangian $L_g$.
Our
investigations are independent of the form of gravitational field
equations
and whether they hold at all, thus we do not take $L_g$ into account.}
(signature is $-+++$) is defined as usual by
\begin{equation}\label{EMTVD}
T_{\mu\nu}(\phi,
g)\equiv-\frac{2}{\sqrt{-g}}\frac{\delta(\sqrt{-g}L)}{\delta
g^{\mu\nu}}.
\end{equation}
As is well known, the definition of variational derivative implies that
the
tensor $T_{\mu\nu}$ is not affected by adding a total divergence to
the Lagrangian, a
property that we shall use in the sequel. In concrete computations,
one obtains the
variational stress tensor  by taking the variation of the action
functional, then
factorizing the variation
$\delta g^{\mu\nu}$ of the metric: this requires dropping a total
divergence to get rid of
terms containing covariant derivatives of $\delta g^{\mu\nu}$. Having
in mind this
standard procedure, in the subsequent computations we rely on the
following
expression, equivalent to (\ref{EMTVD}):
\begin{equation}\label{EMT}
\delta_g(\sqrt{-g}L)= -\frac{1}{2}\sqrt{-g}\left[T_{\mu\nu}(\phi, g)
\delta g^{\mu\nu}+\hbox{div}\right],
\end{equation}
where div means a full divergence which we will sometimes omit, as this term
plays no role in our discussion;
we will mark its presence from time to time to display an exact
equality. In evaluating the variation in eq.~(\ref{EMT}) one assumes
that
$\phi$ is a fundamental field, i.e. is {\em not\/} affected by metric
variations, $\delta_g\phi=0$. This is the case of the vector potential
(one--form) $A_{\mu}$ in electrodynamics while $A^{\mu}$ is already
metric dependent with $\delta_g A^{\mu}=A_{\nu}\delta g^{\mu\nu}$.
Hence, in evaluating $\delta_g L$ one takes into account only the
explicit dependence of $L$ on $g_{\mu\nu}$ and $g_{\mu\nu,\alpha}$
(or covariantly, on $g_{\mu\nu}$ and $\Gamma^{\alpha}_{\mu\nu}$). \\

In terms of the new field $\phi'$ and of the transformed Lagrangian 
$L'$, the stress tensor
of the theory is re-expressed as follows:
\begin{equation}\label{NEMT}
\delta_g(\sqrt{-g}L'(\phi', \xi, g))\equiv
-\frac{1}{2}\sqrt{-g}\left[T'_
{\mu\nu}(\phi', \xi, g)
\delta g^{\mu\nu}+\hbox{div}\right].
\end{equation}
To evaluate $T'_{\mu\nu}(\phi', \xi, g)$ one assumes that the
appropriate
(covariant or contravariant) components of the field $\xi$ are metric
independent, i.e. $\delta_g\xi=0$, while metric variations of the new
dynamical field $\phi'$ are determined by the inverse transformation
$\phi'=\varphi^{-1}(\phi, \xi, g)$, i.e.
\begin{equation}\label{deltaphi}
\delta_{\varphi}\phi'=\frac{\partial \varphi^{-1}}{\partial g^{\mu\nu}}
\delta g^{\mu\nu}+
\frac{\partial \varphi^{-1}}{\partial g^{\mu\nu}{}_{,\alpha}}
\delta g^{\mu\nu}{}_{,\alpha}.
\end{equation}
We denote this variation by $\delta_{\varphi}\phi'$ to emphasize that
$\phi'$
and $g_{\mu\nu}$ are {\em not\/} independent fields: the value
$\phi'(p)$
at any point $p$ depends both on $\phi(p)$ and $g_{\mu\nu}(p)$. Any
scalar or tensor function $f(\phi', \nabla \phi', g)$ depends on the
metric both explicitly (including the connection $\Gamma$) and
implicitly
via $\phi'$, therefore its metric variation is determined by the
{\em substantial ({\rm or} total) variation\/} $\overline{\delta}_g$,
\begin{equation}\label{5}
\overline{\delta}_g f\equiv \delta_g f+\delta_{\varphi}f.
\end{equation}
Here $\delta_g f$ is the variation taking into
account only the explicit metric dependence of $f$, i.e.
\begin{equation}\label{6}
\delta_g f\equiv\frac{\partial f}{\partial g^{\mu\nu}}
\delta g^{\mu\nu}+
\frac{\partial f}{\partial \nabla\phi'}
\delta_g\nabla\phi'.
\end{equation}
To compute $\delta_g\nabla\phi'$ one writes symbolically $\nabla\phi'=
\partial\phi'-\phi'\Gamma$ (assuming that $\phi$ is a covariant tensor)
and recalling that $\delta_g$ does not act on $\phi'$ one gets
\begin{equation}
\delta_g\nabla\phi'=-\phi'\delta\Gamma.
\end{equation}
On the other hand the variation $\delta_{\varphi}$ takes
into account the metric dependence of $f$ via
$\phi'=\varphi^{-1}(\phi, \xi, g)$, then
\begin{equation}\label{8}
\delta_{\varphi}f\equiv \frac{\partial f}{\partial \phi'}
\delta_{\varphi}\phi'+\frac{\partial f}{\partial \nabla\phi'}
\delta_{\varphi}\nabla\phi'
\end{equation}
with $\delta_{\varphi}\phi'$ given by (\ref{deltaphi}) and
\begin{eqnarray}\label{9}
\delta_{\varphi}\nabla\phi'&=&\delta_{\varphi}(\partial \phi'-\Gamma
\phi')=\partial \delta_{\varphi}\phi'-\Gamma\delta_{\varphi}\phi'
\nonumber\\
&=&\nabla\delta_{\varphi}\phi',
\end{eqnarray}
thus $\delta_{\varphi}$ commutes with the covariant derivative $\nabla$.
For a function $f(\phi, \nabla \phi, g)$ the operators
$\overline{\delta}_g$
and $\delta_g$ coincide, i.e.
\begin{eqnarray}\label{10}
\overline{\delta}_g f(\phi, \nabla \phi, g)&=&\delta_g f(\phi, \nabla
\phi, g)
\nonumber\\
&=& \frac{\partial f}{\partial g^{\mu\nu}}
\delta g^{\mu\nu}+
\frac{\partial f}{\partial \nabla\phi}\delta_g\nabla\phi,
\end{eqnarray}
where
\begin{equation}
\delta_g\nabla\phi=\delta_g(\partial\phi-\Gamma\phi)= -\phi\delta\Gamma.
\end{equation}
Accordingly, $\delta_g$ on the l.h.s. of eq. (\ref{NEMT}) should be
replaced by
$\overline{\delta}_g$. \\

The identity (\ref{NL}), which is valid for {\em all\/} $\phi$,
$g_{\mu\nu}$
and transformations $\varphi$, implies
\begin{equation}
\delta_g L(\phi, g)=\overline{\delta}_g L'(\phi',\xi, g),
\end{equation}
what in turn implies the crucial equality
\begin{equation}\label{ovv}
T_{\mu\nu}(\phi,g)=T'_{\mu\nu}(\phi',\xi,g);
\end{equation}
in general the two tensors depend differently on their arguments. We 
stress that
in this way we have {\em not} constructed the stress tensor for the 
{\em different} theory
(also described by the Lagrangian $L'$) in which $\phi'$ would represent a new,
metric--independent field variable; this tensor would {\em not} 
coincide with $T_{\mu\nu}$
unless the fields $\phi$ and $\phi'$, related by the transformation 
$\varphi$, are also {\em
solutions} of the respective field equations: an expression for this 
new stress tensor can be
found, for instance, in \cite{PP} (notice that their equation (2.11) 
holds whenever
the transformation does not depend on the derivatives of the metric, 
otherwise an additional
term occurs). Here, we are dealing with field
redefinitions {\em within} the same theory, not with transformations 
relating theories which
are dynamically equivalent but different in their physical 
interpretation\footnote{As far as
one takes the variation of
$L'$ with respect to $\phi'$ to derive the field equations, it is 
irrelevant to state
whether this field is metric independent or not: the resulting 
equations are the same, and
this may be the reason why the distinction between the two viewpoints 
is usually not
remarked in the literature. However, the variational derivatives of 
$L'$ w.~r.~to
$g^{\mu\nu}$ are different in the two cases.}.

We are interested first in finding out a generic relationship between
$T_{\mu\nu}(\phi,g)$ and $T_{\mu\nu}(\phi',g)$ and then in its reduced
version in the case of a metric--dependent symmetry transformation
$\phi=\varphi(\phi', \xi, g)$ with a specific $\varphi$. To this end
we explicitly evaluate
$T'_{\mu\nu}(\phi',\xi,g)$ from the definition (\ref{NEMT}) and then apply
the
equality (\ref{ovv}). \\
It is convenient to write the transformed Lagrangian as a sum
\begin{equation}\label{deltalag}
L'(\phi', \xi, g)\equiv L(\phi', g)+\Delta L'(\phi', \xi, g),
\end{equation}
this is a definition of $\Delta L'(\phi', \xi, g)$. This splitting
allows one to obtain $T_{\mu\nu}(\phi',g)$ upon applying
$\overline{\delta}_g$ to eq.~(\ref{deltalag}). It is furthermore
convenient to
make the inverse transformation in the term $\Delta L'$, then
\begin{equation}
\Delta L'(\phi', \xi, g)=\Delta L'(\varphi^{-1}(\phi, \xi, g),\xi,g)
\equiv \Delta L(\phi, \xi, g),
\end{equation}
and this is a definition of $\Delta L(\phi, \xi, g)$. Then
eq.~(\ref{deltalag})
takes the form
\begin{equation}\label{deltal}
L'(\phi', \xi, g)= L(\phi', g)+\Delta L(\phi, \xi, g).
\end{equation}

The Lagrange equations for $\phi$ arising from $L(\phi,g)$ are
\begin{equation}\label{17}
E(\phi)\equiv \frac{\delta L}{\delta \phi}=\frac{\partial L}{\partial
\phi}-
\nabla \left(\frac{\partial L}{\partial \nabla \phi}\right)=0;
\end{equation}
the tensor $E(\phi)$ has the same rank and
symmetry as the field $\phi$,
and the definition (\ref{EMT}) provides the energy--momentum tensor for
$\phi$,
\begin{equation}\label{18}
T_{\mu\nu}(\phi,g)\delta g^{\mu\nu}=\delta g^{\mu\nu}(g_{\mu\nu}L -
2\frac{\partial L}{\partial g^{\mu\nu}})- 2\frac{\partial
L}{\partial\nabla\phi}\delta_g\nabla\phi.
\end{equation}
We can now evaluate $T'_{\mu\nu}(\phi',\xi,g)$:
\begin{eqnarray*}
\overline{\delta}_g (\sqrt{-g}L'(\phi', \xi, g))&=&-\frac{1}{2}
\sqrt{-g}[g_{\mu\nu}\delta g^{\mu\nu}(L(\phi', g)+\Delta L(\phi, \xi,
g))
\\
&&
-2\overline{\delta}_g (L(\phi', g)+\Delta L)].
\end{eqnarray*}
Here from eqs.~(\ref{5}), (\ref{6}) and (\ref{8})
$$
\overline{\delta}_g L(\phi', g)=\left[\frac{\partial
L(\phi')}{\partial g^{\mu\nu}}\delta
g^{\mu\nu}+ \frac{\partial L(\phi')}{\partial \nabla\phi'}
\delta_g\nabla\phi'\right]
+\left[\frac{\partial L(\phi')}{\partial \phi'}
\delta_{\varphi}\phi'+\frac{\partial L(\phi')}{\partial \nabla\phi'}
\delta_{\varphi}\nabla\phi'\right],
$$
and the first square bracket contributes to $T_{\mu\nu}(\phi',g)$
while the second one is equal, after applying (\ref{9}), to
\begin{equation}
\left[\frac{\partial L(\phi')}{\partial \phi'}
\delta_{\varphi}\phi'-\delta_{\varphi}\phi'\nabla
\left(\frac{\partial L(\phi')}{\partial
\nabla\phi'}\right)+\hbox{div}\right].
\end{equation}
Employing (\ref{10}) one has $\overline{\delta}_g\Delta L(\phi, \xi,g)=
\delta_g\Delta L$. Then employing eqs.~(\ref{18}), (\ref{17}) and
(\ref{ovv}) and dropping
a full divergence one arrives at the fundamental relationship
\begin{eqnarray}\label{identity}
\delta g^{\mu\nu}[T_{\mu\nu}(\phi',g)-T_{\mu\nu}(\phi,g)+g_{\mu\nu}
\Delta L(\phi, \xi, g)]
\nonumber\\
-2E(\phi')\delta_{\varphi}\phi'-2\delta_g\Delta L(\phi, \xi,g)&=&0.
\end{eqnarray}
This is an identity (up to a total divergence) valid for any field,
Lagrangian and any
field  transformation. The two terms containing $\Delta L$ can be
written
together as
\begin{equation}
-\frac{2}{\sqrt{-g}}\delta_g(\sqrt{-g}\Delta L).
\end{equation}
We remark that if all full divergence terms were kept in the derivation
of the identity, a divergence term would replace zero on the r.h.s. of
eq. (21). However a total divergence cannot cancel the last three
terms on the l.h.s. of the identity since in general these terms
do not sum up into a divergence. Therefore the difference
$T_{\mu\nu}(\phi',g)-T_{\mu\nu}(\phi, g)$ does not vanish
in general. \\

The transformation $\phi\mapsto\phi'$ is a \emph{symmetry
transformation} of the theory (of the Lagrangian) iff
$L'(\phi',\xi, g)=L(\phi',g)+\hbox{div}$, i.e. if
   $\Delta L(\phi, \xi,g)=\hbox{div}$ or is zero. According to the
proposition ``the metric variation of a divergence is another
divergence", adding a covariant divergence to $L(\phi,g)$
does not affect the variational stress tensor; in a similar way one
shows that the variation with respect to the dynamical field $\phi$
of a full divergence gives rise to another divergence, thus the
Lagrange field equations remain unaffected too. The equality
$L'(\phi',\xi, g)=L(\phi',g)+\hbox{div}$ should hold identically for
a symmetry independently of whether the field equations are
satisfied or not. It is worth stressing that we impose no
restrictions on the transformation
$\varphi$ and on possible symmetries --- they should only
continuously (actually
smoothly) depend on a set of parameters corresponding to a set of
scalar functions or to the
components of a vector or tensor field $\xi$; discrete
transformations, like reflections, are
excluded. The identity (\ref{identity}) has deeper consequences
usually when the
transformation
$\varphi$ depends on the spacetime metric (possibly through covariant
derivatives
of the field $\xi$). For any symmetry the  identity
(\ref{identity}) reduces to
\begin{equation}\label{gaugid}
\delta g^{\mu\nu}[T_{\mu\nu}(\phi',g)-T_{\mu\nu}(\phi,g)]
-2E(\phi')\delta_{\varphi}\phi'=0
\end{equation}
since for $\Delta L=\hbox{div}\equiv \nabla_{\alpha}A^{\alpha}
(\phi, \xi,g)$ one has
\begin{equation}
-\frac{2}{\sqrt{-g}}\delta_g(\sqrt{-g}\Delta L)=-2\nabla_{\alpha}
(\delta_g A^{\alpha}-\frac{1}{2}A^{\alpha}g_{\mu\nu}\delta g^{\mu\nu})
\end{equation}
and the divergence is discarded. As a trivial example, consider
Maxwell electrodynamics. Here
$\phi=A_{\mu}$, $\phi'= A_{\mu}+\partial_{\mu}f$ with arbitrary $f$;
since the gauge
transformation is metric independent, $\delta_{\varphi}\phi'=0$. Then the
term
$E(\phi')\delta_{\varphi}\phi'=\nabla_\nu
F^{\mu\nu}\delta_{\varphi}\phi'$ vanishes giving
rise to the gauge invariance of $T_{\mu\nu}$ independently of Maxwell
equations. A nontrivial
example is considered in the next section.

Since the term $E(\phi')\delta_{\varphi}\phi'$ is different from zero in
general (i.e. for fields not being solutions and for metric--dependent
symmetry
transformations), one arrives at the  conclusion: \emph{the metric
energy--momentum
tensor for a theory  having a symmetry does not possess this
symmetry}. The symmetry implies that
\begin{equation}
L(\phi,g)=L(\phi',g)+\hbox{div},
\end{equation}
i.e.~replacing $\phi$ by $\phi'$ in the Lagrangian alters the action
\begin{equation}
S[\phi,g]=\int d^4x \sqrt{-g}L(\phi,g)
\end{equation}
by at most a surface term. Yet the associated integral
\begin{equation}
\delta_g S=-\frac{1}{2} \int d^4x\sqrt{-g}T_{\mu\nu}(\phi, g)
\delta g^{\mu\nu}
\end{equation}
does depend on the transformation. It is only
{\em for solutions\/}, $E(\phi')=E(\phi)=0$, that the energy--momentum
tensor does possess the same symmetry,
$T_{\mu\nu}(\phi',g)=T_{\mu\nu}(\phi,g)$. \\
In physics one is mainly interested in quantities built up of
solutions of equations of motion, but from the mathematical
viewpoint it is worth noticing that the symmetry property is {\em not\/}
carried over from $S$ to $\delta_g S$.

\section{Linear massless spin--2 field in Minkowski space}
In gauge theories of particle physics the field potentials are
exterior forms since the fields
carry spin one. Then the gauge transformations are independent of the
spacetime metric and the
identity (\ref{gaugid}) implies gauge invariance of the stress tensor
for arbitrary fields,
not necessarily being solutions to the field equations. Yet it is
characteristic for gauge
theories that \emph{for integer spins larger than one} a gauge
transformation necessarily
involves covariant derivatives of vector or tensor fields \cite{FWF},
giving rise to gauge dependent \emt{s}. However the gauge symmetry in
these theories holds only
in flat spacetime (more precisely, the Lagrangians are gauge
invariant only in Minkowski
space, while Lagrange field equations are gauge invariant in empty
spacetimes, i.e.~for
$R_{\mu\nu}=0$) and disappears in a generic curved one. Yet the
stress tensors are gauge
dependent for solutions even in
Minkowski space. This case is comprised in the generic theory
developed in the previous section, identity (\ref{identity}), but is
rather misleading and confusing and because of its relevance for
gravity it deserves a
separate treatment. \\

We illustrate the effect in the case of spin--two field $\psi_{\mu\nu}=
\psi_{\nu\mu}$. Any Lagrangian $L(\psi_{\mu\nu}, \psi_{\mu\nu;\alpha},
g_{\alpha\beta})$ generates equations of motion of the form
\begin{equation}
E^{\mu\nu}(\psi)\equiv {\delta L\over \delta \psi_{\mu\nu}}=
{\partial L\over \partial \psi_{\mu\nu}}-
\nabla_{\alpha} ({\partial L\over \partial\psi_{\mu\nu;\alpha}})=0
\end{equation}
and the expression (\ref{18}) for the stress tensor reads now
\begin{eqnarray}
T_{\mu\nu}(\psi,g)&=&g_{\mu\nu}L -
2{\partial L\over \partial g^{\mu\nu}}+
\nonumber\\
&&
2\nabla_{\sigma}[2{\partial L \over
\partial\psi_{\beta(\alpha;\sigma)}}\psi_{\beta(\mu}g_{\nu)\alpha}-
{\partial L \over
\partial\psi_{\beta(\alpha;\tau)}}\psi^{\sigma}{}_{\beta}g_
{\alpha(\mu}g_{\nu)\tau}].
\end{eqnarray}
Let $L$ be gauge--invariant under $\psi_{\mu\nu}\mapsto
\psi'_{\mu\nu}\equiv \psi_{\mu\nu}+\xi_{\mu;\nu}+\xi_{\nu;\mu}$
in {\em flat spacetime\/}. Using covariant expressions in Minkowski
space (coordinates need not be Cartesian ones) this symmetry implies
that
\begin{equation}
L(\psi,g)=L(\psi',g)+\nabla_{\alpha}A^{\alpha}(\psi,\xi, g)
\end{equation}
for some vector $A^{\alpha}$ {\em provided that\/} the covariant
derivatives commute, $\nabla_{\mu}\nabla_{\nu}=\nabla_{\nu}\nabla_{\mu}$.
In
a generic curved spacetime the
"gauge" transformation is no more a symmetry since
\begin{equation}
L(\psi,g)=L(\psi',g)+{\rm div}+\Delta L(\psi,\xi,g),
\end{equation}
with div denoting a full divergence while $\Delta L$ is a sum of terms
proportional to Riemann tensor. For $\delta_g \psi_{\mu\nu} = 0 =
\delta_g \xi_{\mu}$
one gets for this gauge transformation
\begin{equation}
\delta_\varphi \psi'_{\mu\nu}=-2 \xi_\alpha\delta\Gamma^\alpha_{\mu\nu}.
\end{equation}
In this case the fourth term in the identity (\ref{identity}) reads
\begin{equation}
-2E^{\mu\nu}(\psi')\delta_\varphi \psi'_{\mu\nu}=4
E^{\mu\nu}(\psi')\xi_\alpha\delta\Gamma^\alpha_{\mu\nu}
\end{equation}
and discarding the full divergence arising from $\delta
g^{\mu\nu}{}_{;\alpha}$ one arrives at
the following explicit form of (\ref{identity}) for the linear
massless spin--2 field
$\psi_{\mu\nu}$,
\begin{eqnarray}\label{spin2id}
\delta g^{\mu\nu}\lbrace
T_{\mu\nu}(\psi',g)-T_{\mu\nu}(\psi,g)+g_{\mu\nu}
\Delta L(\psi, \xi, g)\nonumber\\
+2\nabla_\alpha[2\xi_{(\mu}E_{\nu)}^\alpha(\psi',g)-\xi^\alpha
E_{\mu\nu}(\psi',g)]\rbrace
\\
-2\delta_g\Delta L(\psi, \xi,g)&=&0.\nonumber
\end{eqnarray}

We recall that this identity is valid only for the transformation
$\psi'_{\mu\nu}=
\psi_{\mu\nu}+2\xi_{(\mu;\nu)}$. One is interested in evaluating this
identity in flat
spacetime, $R_{\alpha\beta\mu\nu}(g)=0$, what will be symbolically
denoted by $g=\eta$.
   One has $\left.\Delta L\right|_{g=\eta}={\rm div}$ while
$\left.\delta_g\Delta L\right|_{g=\eta}\not=0$. In fact, $\Delta L$
is a sum of terms of the
form $t^{\alpha\beta\mu\nu}R_{\alpha\beta\mu\nu}$ where
$t^{\alpha\beta\mu\nu}$ is made up of
$\psi_{\mu\nu}$, $\psi_{\mu\nu;\alpha}$, $\xi_{\alpha}$,
$\xi_{\alpha;\mu}$ and
$\xi_{\alpha;\mu\nu}$ (see eq.~(\ref{sotto}) below). Then
$\left.\delta_g\left(t^{\alpha\beta\mu\nu}R_{\alpha\beta\mu\nu}\right)
\right|_{g=\eta}=
0+\left.t^{\alpha\beta\mu\nu}\delta
R_{\alpha\beta\mu\nu}\right|_{g=\eta}$ and the latter term
does not vanish, in general, even for $R_{\alpha\beta\mu\nu}(g)=0$.

Let us denote the
expression in square backets in (\ref{spin2id}) by
$F^\alpha_{\mu\nu}(\psi',\xi,g)$. In flat
spacetime the gauge invariance implies (for any $\psi_{\mu\nu}$, not
necessarily a solution)
that $E_{\mu\nu}(\psi',\eta)=E_{\mu\nu}(\psi,\eta)$ and then
$\left.\nabla_\alpha
F^\alpha_{\mu\nu}(\psi',\xi,g)\right|_{g=\eta}=\nabla_\alpha
F^\alpha_{\mu\nu}(\psi,\xi,\eta)$. Assuming that $\psi_{\mu\nu}$ is a
solution in Minkowski
space, $E_{\mu\nu}(\psi,\eta)=0$, one gets that $\nabla_\alpha
F^\alpha_{\mu\nu}(\psi,\xi,\eta)=0$. Thus, the indentity
(\ref{spin2id}) reduces for
solutions of the fields equations, in Minkowski space, to
\begin{equation}\label{spin2sol}
\delta g^{\mu\nu}\lbrace
T_{\mu\nu}(\psi',\eta)-T_{\mu\nu}(\psi,\eta)\rbrace
-2\left.\delta_g\Delta L(\psi, \xi,g)\right|_{g=\eta}=0.
\end{equation}

This relationship (not an identity) shows that the stress tensor is
\emph{not} gauge invariant
even in flat spacetime. In other terms, the symmetry properties of
the Lagrangian in Minkowski
space are insufficient for determining symmetry properties of the
metric stress tensor in this
spacetime. To this end one must investigate the Lagrangian in a
general curved spacetime and
its behaviour there is relevant for the stress tensor in flat spacetime.

The gauge invariant linear massless spin--2 field was introduced by
Fierz and Pauli \cite{FP}
and its \lag\ usually bears their names. Finding out the appropriate
\lag\ is not
straightforward and they had to use a rather indirect procedure.  For
our
purposes it is adequate to employ the dynamical equivalence of this
field to linearized \GR\
and following Aragone and Deser \cite{AD1} generate a \lag\ for it as
the \lag\ for a metric
perturbation around Minkowski space.
The resulting \lag\ is of first order and may be unambiguously
expressed in any curved
background as
\begin{equation}\label{lw}
L_W(\psi,g)=\frac{1}{4}\left(-\psi^{\mu\nu;\alpha}\psi_{\mu\nu;\alpha}
+2\psi^{\mu\nu;\alpha}\psi_{\alpha\mu;\nu}
-2\psi^{\mu\nu}{}_{;\nu}\psi_{;\mu}+\psi^{;\mu}\psi_{;\mu}\right)
\end{equation}
with $\psi=g^{\mu\nu}\psi_{\mu\nu}$. This \lag\ appeared first in the
textbook \cite{W} and
will be referred hereafter to as \emph{Wentzel \lag}. Actually in
Minkowski space the choice
of a
\lag\ for $\psi_{\mu\nu}$ is not unique and a number of equivalent
\lag{s} exist\footnote{In
fact the requirement of gauge invariance uniquely fixes a quadratic
Lagrangian for the field
up to some boundary terms \cite{HOR}.}. For example
one can replace the second term in (\ref{lw}) by a more symmetric term
$\psi^{\mu\nu}{}_{;\nu}\psi_{\mu\alpha}{}^{;\alpha}$ and the
resulting \lag\ $L_S$ differs from
$L_W$ by a full divergence. However in a curved spacetime the two
\lag{s} differ by
a curvature term,
$L_S(\psi,g)=L_W+\mathrm{div}+H$,
where $H\equiv \psi^{\nu}{}_{\mu}\psi^{\mu\alpha}{}_{;[\nu\alpha]}=
\frac{1}{2}\psi^{\alpha\beta}(\psi_{\beta}{}^{\mu} R_{\mu\alpha} +
\psi^{\mu\nu}R_{\mu\alpha\beta\nu})$. There exists also a second--order
\lag
\begin{equation}\label{lii}
L_{I\! I}(\psi,g)=-\frac{1}{2}\psi^{\mu\nu}\Glin_{\mu\nu}(\psi,g)
\end{equation}
where
\begin{eqnarray}\label{glin}
\Glin_{\mu\nu}(\psi,g)&\equiv&\frac{1}{4}\left(-\psi_{\mu\nu;\alpha}
{}^{;\alpha}
+\psi^{\alpha}{}_{\mu;\nu\alpha}+\psi^{\alpha}{}_{\nu;\mu\alpha}
-\psi_{;\mu\nu}\right.\nonumber\\
&&\left.-g_{\mu\nu}\psi_{\alpha\beta}{}^{;\alpha\beta} +
g_{\mu\nu}\psi_{;\alpha}{}^{;\alpha}\right)
\end{eqnarray}
is the Einstein tensor linearized around Minkowski space and then
formally written down for an
arbitrary background (actually, if $G_{\mu\nu}$ is linearized around
a curved background, then
there appear in the expansion additional terms depending on the
background curvature). This
\lag\ is equivalent (also in a curved spacetime) to Wentzel \lag,
$L_{I\!
I}(\psi,g)=L_{W}(\psi,g)+\mathrm{div}$.

The Lagrange equations of motion arising from $L_{W}$ (or
equivalently from $L_{I\! I}$) in
any spacetime are
\begin{equation}\label{eqs}
E_{\mu\nu}(\psi,g)=-\Glin_{\mu\nu}(\psi,g)=0,
\end{equation}
while those resulting from $L_S$ contain additional curvature terms.
However, as is well known
(\cite{AD2},\cite{AD1}), the linear spin--2 field is inconsistent in
a curved spacetime since
both sets of field equations develop additional constraints due to
the curvature. We shall
assume that eqs.~(\ref{eqs}) hold only in flat spacetime.

The inequivalence of $L_W$ and $L_S$ (and most other Lagrangians) raises
the
problem of which \lag\ should be used, since
the stress tensors associated with these \lag{s} will be different
even in flat spacetime.
The inconsistency of the theory in the presence of a gravitational
field shows that a proper
choice does not exist. Equivalence of $L_W$ and $L_{I\! I}$ points
rather to $L_W$ and we
shall employ Wentzel \lag. Actually the problem of gauge dependence
of any stress tensor is
independent of the choice; in each case computations are of the same
length and give the same
outcome.

The metric \emt\ generated by Wentzel \lag\ is
\begin{eqnarray}\label{tw}
T_{\mu\nu}^{W}(\psi,\eta)&=&
2\psi_{\alpha(\mu}\psi_{\nu)\beta}{}^{;\alpha\beta}
-\psi_{\mu\nu;\alpha\beta}\psi^{\alpha\beta}
-\psi_{\alpha(\mu}\psi_{;\nu)}{}^{;\alpha}
+\frac{1}{2}\psi_{\mu\nu}\Box\psi \nonumber\\ &&
-\psi_{\mu\nu}\psi^{\alpha\beta}{}_{;\alpha\beta}
+\frac{1}{2}g_{\mu\nu}\psi^{\alpha\beta}\psi_{;\alpha\beta}
-2\psi_{\alpha\beta;(\mu}\psi_{\nu)}{}^{\alpha;\beta}
+\frac{1}{2}\psi_{\alpha\beta;\mu}\psi^{\alpha\beta}{}_{;\nu} \nonumber\\
&&
+2\psi_{(\mu}{}^{\alpha;\beta}\psi_{\nu)(\alpha;\beta)}
-\psi_{\mu\nu;\alpha}\psi^{\alpha\beta}{}_{;\beta}
-\frac{1}{2}\psi_{\mu\nu;\alpha}\psi^{;\alpha}
+\psi_{;(\mu}\psi_{\nu)\alpha}{}^{;\alpha} \nonumber\\ &&
-\frac{1}{2}\psi_{;\mu}\psi_{;\nu}
+\frac{1}{4}g_{\mu\nu}\left(-\psi^{\alpha\beta;\sigma}\psi_{\alpha\beta;
\sigma}
+2\psi^{\alpha\beta;\sigma}\psi_{\sigma\alpha;\beta}
+\psi^{;\alpha}\psi_{;\alpha}
\right),
\end{eqnarray}
where $\Box\psi\equiv\nabla^\alpha\nabla_\alpha\psi$. This covariant
expression holds only in
flat spacetime since in deriving it one assumes that the covariant
derivatives commute.

Now one can prove the gauge dependence of $T_{\mu\nu}^{W}(\psi,\eta)$
either directly from its
explicit form (\ref{tw}) or employing the relationship
(\ref{spin2sol}). The latter method is
more convenient if one wishes to show (as is done in the next
section) that this deficiency is
not peculiar to $T_{\mu\nu}^{W}$ but is a generic feature of all
\lag{s} which are equivalent
to $L_W$ in flat spacetime.

Wentzel \lag\ and any other \lag\ equivalent to it is gauge invariant
only in flat spacetime.
In fact, writing (\ref{deltal}) as
\begin{equation}
L_W(\psi, g)= L_W(\psi', g)+\Delta L_W(\psi, \xi, g).
\end{equation}
one finds (disregarding a divergence) that
\begin{eqnarray}\label{sotto}
\Delta L_W(\psi, \xi,
g)&=&R_{\alpha\beta\mu\nu}(h^{\mu\beta;\alpha}\xi^{\nu}-h^{\nu\beta}
\xi^{\mu;\alpha})\nonumber\\
&&+R_{\alpha\beta}(h^{\alpha\mu}\xi^{\beta}{}_{;\mu}-h^{\alpha\mu}{}_{;\mu
}\xi^{\beta}
+2\xi^{\alpha}{}_{;\mu}\xi^{(\beta;\mu)}-\xi^{\alpha;\beta}\xi^{\mu}{}_{;
\mu})
\nonumber\\
&&+R_{\mu\nu}(\xi_{\alpha}{}^{;\alpha\mu}\xi^{\nu}-\xi^{\mu;\nu\alpha}
\xi_{\alpha}),
\end{eqnarray}
where $h_{\mu\nu}\equiv\psi_{\mu\nu}-\frac{1}{2}g_{\mu\nu}\psi$.

One notes in passing that the tensor $E_{\mu\nu}$ is gauge invariant
in any empty spacetime
$(R_{\mu\nu}=0)$ \cite{AD2},
\begin{eqnarray}
E_{\mu\nu}(\psi',
g)&=&E_{\mu\nu}(\psi, g)-\xi^{\alpha}R_{\mu\nu;\alpha}
-2\xi_{\alpha;(\mu}R_{\nu)}{}^{\alpha}+\nonumber\\
&&+g_{\mu\nu}\left(\frac{1}{2}\xi^{\alpha}R_{;\alpha}
+\xi^{\alpha;\beta}R_{\alpha\beta}\right),
\end{eqnarray}
i.e.~even in the backgrounds where the gauge symmetry of $L_W$ is
broken.

The variation $\delta_g\Delta L_W$ evaluated in Minkowski space is
different from zero,
showing the gauge dependence of $T_{\mu\nu}^{W}$. The expression is
extremely involved; in the
special case of the solution $\psi_{\mu\nu}=0$ it is shown in
eq.~(\ref{nove}) in the next
section.

\section{Nonexistence of a gauge invariant metric stress tensor and
the problem of
gravitational energy density}

The fact that the stress tensor $T_{\mu\nu}^{W}$ depends on the gauge
was known long ago
\cite{AD2} (actually Aragone and Deser investigated a \lag\ different
from $L_W$, giving rise
to a stress tensor different from $T_{\mu\nu}^{W}$, nevertheless the
conclusion was clearly
the same). More interesting is the problem whether there exists a
\lag\ $L_K(\psi,g)$, which is
equivalent to $L_W$ in Minkowski space, but which generates a
different, gauge--invariant
stress tensor $T_{\mu\nu}^{K}$ in this spacetime. The no--go theorem
stating that such gauge
invariant stress tensor does not
exist was given in \cite{DMC}. The authors of that work did not
publish a detailed
proof and only referred to the underlying "folk" wisdom. According to
S.~Deser,
for all gauge fields (with metric dependent gauge transformations) in
flat
spacetime the manifest covariance of energy--momentum density objects is
incompatible with their gauge invariance, i.e. these objects are
either covariant
or gauge invariant but not both \cite{D}. In fact, one can always remove
(in a
non--covariant way) the non--physical components of the fields, so
that the result
will have no residual gauge dependence; notice however that this is
not the same as
producing a gauge--independent definition in the usual sense.
Here we give an alternative direct proof of the no--go
theorem based on the relationship (\ref{spin2sol}).

We set by definition
\begin{equation}
L_K(\psi,g)=L_W(\psi,g)+K(\psi,g);
\end{equation}
the term $K$ should reduce to a full divergence in flat spacetime,
while in a curved one it
should contain a sum of terms proportional to the Riemann tensor:
\begin{eqnarray}\label{kappa}
K(\psi,g)&=&\nabla_{\mu}A^{\mu}+a_1
R_{\alpha\beta\mu\nu}\psi^{\alpha\mu}\psi^{\beta\nu}
+a_2 R_{\alpha\beta}\psi^{\alpha\beta}\psi
+a_3 R_{\alpha\beta}\psi^{\alpha\mu}\psi^{\beta}{}_{\mu}\nonumber\\&&
+a_4 R\psi^{\alpha\beta}\psi_{\alpha\beta}
+a_5 R\psi^2.
\end{eqnarray}
The curvature terms in $K$ generate a nonvanishing contribution to
$T_{\mu\nu}^{K}(\psi,\eta)$,
possibly making it gauge independent for an appropriate choice of the
constant coefficients
$a_1,\ldots,a_5$. On the other hand, $L_K$ and $L_W$ give rise to the
same Lagrange field
equations in Minkowski space,
$E_{\mu\nu}(\psi,\eta)=-\Glin_{\mu\nu}(\psi,\eta)=0$.

The expression (\ref{kappa}) is the most general one providing a
physically acceptable stress tensor. In fact, we assume that
\begin{enumerate}
\item $T_{\mu\nu}^{K}$ should be exactly quadratic in $\psi_{\mu\nu}$. In
fact,
the term $K(\psi,g)$ is expected to cancel out the quadratic
gauge--dependent
terms in the stress tensor $T_{\mu\nu}^{W}$: if
$K(\psi,g)$ contains cubic or linear terms, they would  produce new
gauge--dependent
terms in the stress tensor which would not be compensated by terms
arising from $L_W$.
\item $T_{\mu\nu}^{K}$ should contain at most second derivatives of
$\psi_{\mu\nu}$. This is a
physical postulate: the order of field derivatives in the \emt\
should not exceed the order of
the Lagrange equations of motion.
\end{enumerate}
The latter assumption implies that $K$ should be linear in the
Riemann tensor, and should
not contain derivatives of the Riemann tensor itself; moreover, it
should not contain
derivatives of the field $\psi_{\mu\nu}$ multiplied by the curvature
components. Otherwise, it
is easy to see that the metric variation of $K$ would necessarily
produce at least third
derivatives of $\psi_{\mu\nu}$. Therefore, one should set
$K(\psi,g)=\nabla_{\mu}A^{\mu}+k^{\alpha\beta\mu\nu}R_{\alpha\beta\mu\nu}$
,
the coefficients
$k^{\alpha\beta\mu\nu}$ being functions of the field $\psi_{\mu\nu}$
not depending on its
derivatives. The first assumption entails that these functions should
be exactly quadratic in
$\psi_{\mu\nu}$. Hence, we conclude that $K$ cannot contain other
terms besides those
occurring in (\ref{kappa}).

We now apply the formula (\ref{spin2sol}) to $L_K(\psi,g)$: for
solutions of the field
equations in Minkowski space, one has
\begin{equation}\label{Ksol}
\delta g^{\mu\nu}\lbrace
T_{\mu\nu}^{K}(\psi',\eta)-T_{\mu\nu}^{K}(\psi,\eta)\rbrace
-2\left.\delta_g\Delta L_K(\psi,\xi,g)\right|_{g=\eta}=0.
\end{equation}
One concludes that the stress tensor $T_{\mu\nu}^{K}(\psi,\eta)$ is
gauge invariant for
solutions if and only if
\begin{equation}\label{Kcond}
\left.\delta_g\Delta L_K\right|_{g=\eta}=\left.\delta_g(\Delta
L_W+\Delta K)\right|_{g=\eta}=0.
\end{equation}
Our aim is to show that eq.~(\ref{Kcond}) cannot hold for arbitrary
$\xi_\alpha$.
Under the gauge transformation $K$ varies by
\begin{eqnarray}\label{deltaK}
\Delta K(\psi,\xi,g)&=&-a_1R_{\alpha\beta\mu\nu}
(2\psi^{\alpha\mu}\xi^{\beta\nu}+\xi^{\alpha\mu}\xi^{\beta\nu})
\nonumber\\&&
-R_{\alpha\beta}[a_2
(\psi^{\alpha\beta}\xi+\psi\xi^{\alpha\beta}+\xi^{\alpha\beta}\xi)
+a_3(2\psi^{\mu(\alpha}\xi^{\beta)}{}_{\mu}+\xi^{\alpha\mu}\xi^{\beta}{}_
{\mu})]\nonumber\\&&
-R[a_4(2\psi^{\alpha\beta}\xi_{\alpha\beta}+\xi^{\alpha\beta}\xi_
{\alpha\beta})
+a_5 (2\psi\xi+\xi^2)],
\end{eqnarray}
where we introduced the abbreviations
$\xi_{\mu\nu}\equiv\xi_{\mu;\nu}+\xi_{\nu;\mu}$ and
$\xi\equiv g^{\mu\nu}\xi_{\mu\nu}=2\xi^{\mu}{}_{;\mu}$.

If a gauge invariant $T_{\mu\nu}^{K}$ exists, eq.~(\ref{Kcond})
should become an identity
with respect to $\xi_{\alpha}$ for any solution $\psi_{\mu\nu}$ of
$E_{\mu\nu}(\psi,\eta)=0$.
We show that this is not the case for the solution $\psi_{\mu\nu}=0$.
Even in this simplest case
the expression for $\delta_g\Delta K$ in flat spacetime is too long
to be presented here. The
expression for $\delta_g\Delta L_W$ is shorter a bit and is worth
exhibiting,
\begin{eqnarray}\label{nove}
\delta_g\Delta L_W(0,\xi,g)\Big|_{g=\eta}&=&
\delta g^{\mu\nu}\big[
\xi_{(\mu}{}^{;\alpha}\Box\xi_{\nu);\alpha}
+\xi^{\alpha}{}_{;(\mu}\Box\xi_{\nu);\alpha}
-2\xi_{(\mu;\nu)\alpha\beta}\xi^{\alpha;\beta}\nonumber\\&&
-\xi_{(\mu;\nu)}\Box\xi^{\alpha}{}_{;\alpha}
+g_{\mu\nu}\xi^{\alpha;\beta}\xi^{\sigma}{}_{;\sigma\alpha\beta}+
\xi_{\mu;\alpha\beta}\xi_{\nu}{}^{;\alpha\beta}\nonumber\\&&
-\xi_{(\mu;\nu)\alpha}\Box\xi^{\alpha}
-2\xi_{(\mu;\nu)\alpha}\xi^{\beta;\alpha}{}_{;\beta}
+\xi^{\alpha}{}_{;\alpha(\mu}\Box\xi_{\nu)}\nonumber\\&&
+\frac{1}{2}g_{\mu\nu}\xi^{\sigma}{}_{;\sigma\alpha}(\Box\xi^{\alpha}
+\xi^{\beta;\alpha}{}_{;\beta})
\big].
\end{eqnarray}
Each term in eq.~(\ref{nove}) should be
separately cancelled by an appropriate term arising from
$\delta_g\Delta K$, what
implies that eq.~(\ref{Kcond}) decomposes into 27 linear algebraic
equations for the
coefficients $a_i$. Among these, there
are 10 linearly independent equations, and they clearly form and
overdetermined system which
admits no solution. This proves that eq.~(\ref{Kcond}) cannot hold
identically for an
arbitrary vector field $\xi_{\alpha}$ and the stress tensor
$T_{\mu\nu}^{K}$ \emph{cannot} be
gauge invariant, for any choice of the coefficients $a_1,\ldots,a_5$.

The same outcome can indeed be obtained by just evaluating the
difference
$T_{\mu\nu}^{K}(\psi',\eta)-T_{\mu\nu}^{K}(\psi,\eta)$ for $K$ given
in (\ref{kappa}). In this
case, however, the computation is much longer. \\

The physical relevance of the linear massless spin--two field
$\psi_{\mu\nu}$ in flat
spacetime stems from the fact that it is dynamically equivalent to
linearized
General Relativity. Hence this field is closely related to the problem
of
gravitational energy density. A
conventional wisdom says that gravitational energy and momentum densities
are
nonmeasurable quantities. Nevertheless since the very advent of \GR\
there have been numerous attempts to construct a local concept of
gravitational energy.
A gravitational \emt\ is highly desirable for a number of reasons.
For instance, it is emphasized in
\cite{BG} that such a genuinely local tensor is required for a
detailed description
of cosmological perturbations
in the early universe. Approaches to the problem in the framework of
General Relativity include the use of a quasilocal energy--momentum tensor
for
a spatially bounded region (see e.g. \cite{BLY}),
various effective energy--momentum tensors which are
conserved and may be invariant with respect to diffeomorphism--induced
gauge
transformations\footnote{A different viewpoint is based on the fact
that geometrical
General Relativity, including the full nonlinear Einstein field
equations, can be
uniquely derived as a consistent self--coupled theory arising from
the free linear massless spin--2 field theory in Minkowski space
\cite{D1},
\cite{BD}, see also \cite{D2}.
In this approach an energy--momentum tensor for the
gravitational field arises as a part of the field equations in the form
of
a Noether current.} \cite{A}, \cite{ABM} and applications of the
Noether approach with
some version of the Belinfante symmetrization method \cite{BCJ}. Our
considerations
here do not apply to these objects. \\
A completely different approach to the problem
is provided by the field theory approach to
gravitation, according to which gravity is just a tensor field
existing in Minkowski space,
which is the spacetime of the physical world (for a historical review
see \cite{B}). In
these theories of gravity the metric energy--momentum tensor again
serves as the most
appropriate local description of energy for the field \cite{BG}. The
best and most
recent version of field theory of gravitation given in \cite{BG}
satisfies the natural
requirement that all viable theories of gravity should dynamically
coincide in the
weak--field approximation with the linearized \GR\, i.e. gravitation
should be
described by the linear field $\psi_{\mu\nu}$. Furthermore the metric
stress tensor
derived in \cite{BG} has a number of nice properties and
according to the authors, their $T^{\mu\nu}$
is the correct \emt\ for the gravitational field. However, while the
linearized Lagrange
equations of their theory are gauge invariant (as being equivalent to
those for
$\psi_{\mu\nu}$), their \emt\ in this approximation shares the defect
of all the
metric stress tensors for $\psi_{\mu\nu}$, i.e. breaks the gauge
symmetry.
Applying a physically undeniable condition
that the \emt\ should have the same gauge invariance as the field
equations, we conclude that also this approach to gravity does not furnish
a
physically acceptable notion of gravitational energy
density\footnote{In an earlier work
\cite{GPP} it is shown that the field--theoretical formulation of
\GR\ in a Minkowski background does not provide a gauge invariant
\emt, but the authors do
not regard this fact as a defect of their approach.}. Of course, this
does not mean that in
linearized \GR\ there are no conserved tensors which are gauge
invariant, symmetric and
quadratic in (second) derivatives of fields, which to some extent may
play the role of energy
density. The famous Bel--Robinson superenergy tensor \cite{BR} is the
best example.

To avoid any misunderstanding, we emphasize that we do not regard the
field--theoretical
approach to gravity as an alternative theory of gravitation which
might in principle replace
\GR\ as an adequate description of reality. If instead one views this
approach as a different
theory of gravity then one may claim that the ``gauge''
transformation actually maps one
solution of field equations to another \emph{physically distinct}
solution. Then the energy
density need not be gauge invariant and measurements of energy may be
used to discriminate
between two physically different solutions related by the ``gauge''
transformation (which
should then be rather called a ``symmetry transformation''). The
gravitational
field of \cite{BG} or the spin--2 field with the Wentzel \lag\ would then
be
measurable quantities rather than gauge potentials. If the
transformation of these fields is
not an internal gauge but corresponds to a change of physical state,
this raises a difficult
problem of finding out a physical interpretation of it. Clearly it is
not a transformation
between reference frames of any kind.

Here we adopt the opposite view that the field theory approach to
gravity is merely an
auxiliary procedure for constructing notions which are hard to define
in the framework of \GR.
It is commonly accepted that in the weak field limit of \GR\ the
spacetime
metric is measurable only in a very restricted sense: if two almost
Cartesian coordinate
systems are related by an infinitesimal translation
$x'^\mu=x^\mu+\xi^\mu$, then no experiment
can tell the difference of their metrics while the curvature tensor
has the same components in
both systems. This implies that all different coordinate systems
connected by this
transformation actually represent the same physical reference frame
and from the physical
viewpoint the transformation is an internal gauge symmetry
\cite{MTWW}. Thus, showing a
mathematical equivalence of the corresponding field equations is
insufficient to
achieve compatibility of an approach to gravity with the linearized
\GR. The weak field gravity should be described by a gauge potential.
In consequence, any
gravitational energy density should be a gauge invariant quantity.

There are numerous no--go theorems in physics and
it is not unlikely that this one will be somehow circumvented as most
of them have been
   and an acceptable notion
of gravitational energy density will be ultimately defined. However,
this notion cannot be
expressed in terms of the metric \emt\ in a \lag\ field theory since
in the weak--field
approximation the latter cannot be gauge invariant.
Therefore we feel that the theorem closes one line of research of
gravitational energy
density. This makes the quest of this notion harder than previously.\\

\section{Conclusions}

Our approach allows one to clarify in full generality the rather
surprising fact
that the metric \emt\ may not
possess a symmetry property of the underlying field \lag.
This feature, first encountered in the case of gauge symmetry of the
massless
linear spin-two field, turns out to be a generic one. This is due to
the presence
of the spacetime metric. The metric stress tensor acquires a symmetry
only if the
Lagrangian has this symmetry in an open neighbourhood in the space of
Lorentzian
metrics; if the symmetry only holds for a specific spacetime (e.g. in
Minkowski
space), the stress tensor cannot inherit it besides exceptional cases
(when the
variational derivative in the last term in eq.~(\ref{identity})
vanishes). Whenever
the symmetry transformation of the field variables depends on the metric,
the
stress tensor acquires the symmetry only for solutions.

As an obvious application we have used our generic method to the case of
gauge
fields with spins larger than one and rederived in a different way
previous
conclusions \cite{DMC}. The case of these high spin fields has been
additionally
obscured by the well
known fact that these theories are dynamically inconsistent in the
presence of
gravitation and this is closely connected with the gauge invariance
breaking in a
curved spacetime. We
believe that our approach sheds new light on the confusing issue of
these fields.

We expect that the generic picture of symmetry breaking for the metric
energy--momentum tensor in Lagrangian field theories presented in this
work
will find further physical applications besides the case of the gauge
invariance of high spin linear fields. However, even in
this case our method employed to the massless spin--2 field allows to
show
in a simple and direct way that the field--theory approach to
gravitational energy
density is actually hopeless. \\

The authors are grateful to
Stanley Deser for his comments regarding his work on the subject and to
   Marco Ferraris for his suggestions concerning the
gauge dependence of the stress tensor for the Wentzel Lagrangian.
The work of L.M.S. was partly supported by KBN grant no.~2P03D01417
and by the INdAM-GNFM
project \emph{``Metodi Geometrici in Fisica Mate\-matica''}.

\end{document}